\documentclass[aps,pre,twocolumn,superscriptaddress,nofootinbib,floatfix]{revtex4}

\usepackage{dcolumn}
\usepackage{amsmath}
\usepackage{amssymb}
\usepackage{amsthm}

\usepackage{graphics}
\usepackage[english]{babel}
\usepackage[utf8]{inputenc}
\usepackage{graphicx}  
\usepackage{latexsym}  % useful for coding complex math
\usepackage{amsfonts} 
\usepackage{color} 
\usepackage{epstopdf}
\usepackage{enumerate}
\usepackage{subfigure}

\begin{document}

\title{Networks with Growth and Preferential Attachment:  Modeling and Applications}

\author{Gabriel G. Piva\footnote{ \url{https://orcid.org/0000-0001-7636-6568} }}  
\email{gabrielgpiva@gmail.com} 
\affiliation{Departamento de F\'{i}sica, Universidade Federal de Lavras, 37200-900 Lavras, MG, Brazil}
\affiliation{ Departamento de Fisica, Pontif\'{i}cia Universidade Cat\'{o}lica do Rio de Janeiro, 22451-900 Rio de Janeiro, RJ, Brazil}

\author{Fabiano L. Ribeiro\footnote{ \url{https://orcid.org/0000-0002-2719-6061}}}
\email{fribeiro@ufla.br}
 \affiliation{Departamento de F\'{i}sica, Universidade Federal de Lavras, 37200-900 Lavras, MG, Brazil}

\author{Ang\'{e}lica S. Mata\footnote{ \url{https://orcid.org/0000-0002-3892-5274}}}
\email{angelica.mata@ufla.br} 
\affiliation{Departamento de F\'{i}sica, Universidade Federal de Lavras, 37200-900 Lavras, MG, Brazil}%\inst{1} % etc
% \thanks is optional - remove next line if not needed%

\begin{abstract}
In this article we presented a brief study of the main network models with growth and preferential attachment.
Such models are interesting because they present several characteristics of real systems.
We started with the classical model proposed by Barab\`{a}si and Albert~\cite{barabasi1999emergence}: nodes are added to 
the network connecting preferably to other nodes that are more connected. We also presented models that 
consider more representative elements from social perspectives, such as
the homophily between the vertices or the fitness that each node has to build 
connections~\cite{bianconi2001competition,homophilic}. Furthermore, we showed a version of these models including 
the Euclidean distance between the nodes as a preferential attachment rule~\cite{soares2005preferential}. Our objective is to investigate the basic properties of these networks 
as distribution of connectivity, degree correlation, shortest path, cluster coefficient and how these 
characteristics are affected by the preferential attachment rules. Finally, we also provided a comparison of these synthetic  networks with real ones. We found that characteristics as homophily, fitness and geographic distance are significant preferential attachment rules to modeling real networks. These rules can change the degree distribution form of these synthetic network models and make them more suitable to model real networks. 
\end{abstract}

%In fact, our analysis of social and technological networks emphasized that they do not always have a power law degree distribution. 
%
%
\maketitle

\section{Introduction}
\label{sec:intro}

\indent Complex systems has become a widely applied area of research because of everything around us can be described by a 
complex network, including social, technological or biological organisms. The growth and the preferential attachment 
considering that a node has higher probability to connect with a other node that already have many edges are famous ingredients~\cite{barabasi1999emergence} to produce a power law degree distribution, frequently used topology to describe real systems.

In general, it has been shown that real networks present a power law degree distribution with $2<\gamma<3$~  \cite{radicchi2015percolation,Newman,nature_communications,caldarelli2007scale,associatividade,barabasi2016network,phonecalls,email,collaboration}. 
However, this is a controversial topic~\cite{broido2019scale,holme2019,barabasi2016network}. In a recent study, Broido and Clauset~\cite{broido2019scale}
investigated nearly 1000 of real networks using statistical tools. They showed evidences that power-law degree structured is not usual to be found in real-life. They evaluated social,  biological,  technological,  transportation,  and  information  networks. 
Their main conclusion is social networks are weakly scale-free while technological and biological networks are strongly scale-free. However, they also found that $51\%$ of the real data set can be classified as some kind of scale-free category. Barabàsi also arguments\footnote{blog post: https://www.barabasilab.com/post/love-is-all-you-need} that real networks, ruled by growth and preferential attachment, have  power law with an exponential cutoff degree distribution.

In this paper, we investigated social and technological real networks and we found that they can be modeled by networks with growth and preferential attachment. To account for more realistic aspects, we considered other concepts in the preferential attachment 
as fitness~\cite{bianconi2001competition}, homophily~\cite{homophilic}, and Euclidean distance between nodes~\cite{soares2005preferential}. 
Indeed, social systems often present these kind of feature's connections~\cite{currarini2016simple,bisgin2010investigating} 
and real-world systems in general are often embedded in Euclidean 
space~\cite{scalefreelattice,Liu2018,Laniado2017,Lengyel2015GeographiesOA}.
We investigated the phone calls~\cite{phonecalls}, collaboration~\cite{collaboration} and e-mails networks~\cite{email}. The first two are social networks because they describe family, friendship and/or  professional interactions while the email network behaves as a technological network. 
We also found that the email network present a more ``scale-free'' behavior in its degree distribuiton while social networks are better described by a $q$-exponential degree distribution, according to the model proposed by Soares and collaborators~\cite{soares2005preferential}.

The paper is divided as follows: The detailed description of networks models with growth and different rules of preferential attachments are found in section~\ref{sec:models}, where we also studied some properties of these networks as degree distribution and assortativity. The main information and results about the networks are summarized in table~(\ref{tab:table_general}). In section \ref{sec:realnets}, we provided a comparison of these synthetic 
networks with real ones. At last, we presented our final considerations in section~\ref{sec:conclu}.

\section{Network Models with Growth and Preferential Attachment}  

\label{sec:models}

A network model has properties similar to real systems. Networks are considered a powerful tools to represent patterns of connections between parts of systems such as Internet, power grid, 
food webs, social networks, etc \cite{caldarelli2007scale,bollobas2003mathematical,dorogovtsev2002evolution}. Some particular metrics properties, like degree distribution, shortest path length, and clustering coefficient have been attracted attention of physics communities. \\ Watts and Strogatz \cite{watts1998collective} shows that real networks is characterized by average shortest path distance between two vertex and large clustering coefficient, describing these properties by \textit{small-world} model.

Based on that, Barabàsi-Albert \cite{barabasi1999emergence} proposed two basics mechanisms that try to better characterize a real network: growth of system, adding new agents and preferential attachment, where a new agent connects preferentially  with most connected nodes already on the network. The web expands with adding of new documents which links with older or well known sites, for instance. The probability that a new node will connect to a node with $k$ links is proportional to $k$, independently of geographic distance. 

However, there are other examples of real networks whose connectivity may depend on the geographic distance between the nodes, as a power grid. In addition to geographic distance, there may be other relevant ingredients to consider when connecting the elements of the system. Social interaction between people have intrinsic characteristics that should be taken into account as for example 
the influence one person has on another and the affinity between them, representing friendship,
familiar or professional ties. 

To model these features, some networks have been studied through over the years. We presented below some of them that consider preferential attachment rules according to the degree (Barabàsi-Albert model~\cite{barabasi1999emergence}), or the fitness of the node to make connections~\cite{bianconi2001competition}, or the homophiy between them~\cite{homophilic}, and finally, according to the euclidean distance between the nodes~\cite{soares2005preferential}.

\subsection{Barabási-Albert Network}
\label{sec:BA}

To explain in a simple way the behavior of technological networks, such as internet, Barabàsi and 
Albert~\cite{barabasi1999emergence} proposed the following model:

\begin{itemize}
\item  The system starts with $m_0$ nodes connected to each other.
\item At each time step, a new node $j$ is entered on the network and it connects to a random node $i$ chosen at random
with probability $\Pi(k_i|j)$ proportional to its degree ($k_i$), which means

\begin{equation}
 \Pi(k_i|j)=\frac{k_i}{\sum_n k_n}
 \label{eq:ProbConec_BA}
\end{equation}
\end{itemize}
where the normalization $\sum_n k_n$ is the sum over all degree $k_n$ of each node $n$ already connected on the network.

\begin{figure}[!htb]
 \centering
  \includegraphics[width=7cm]{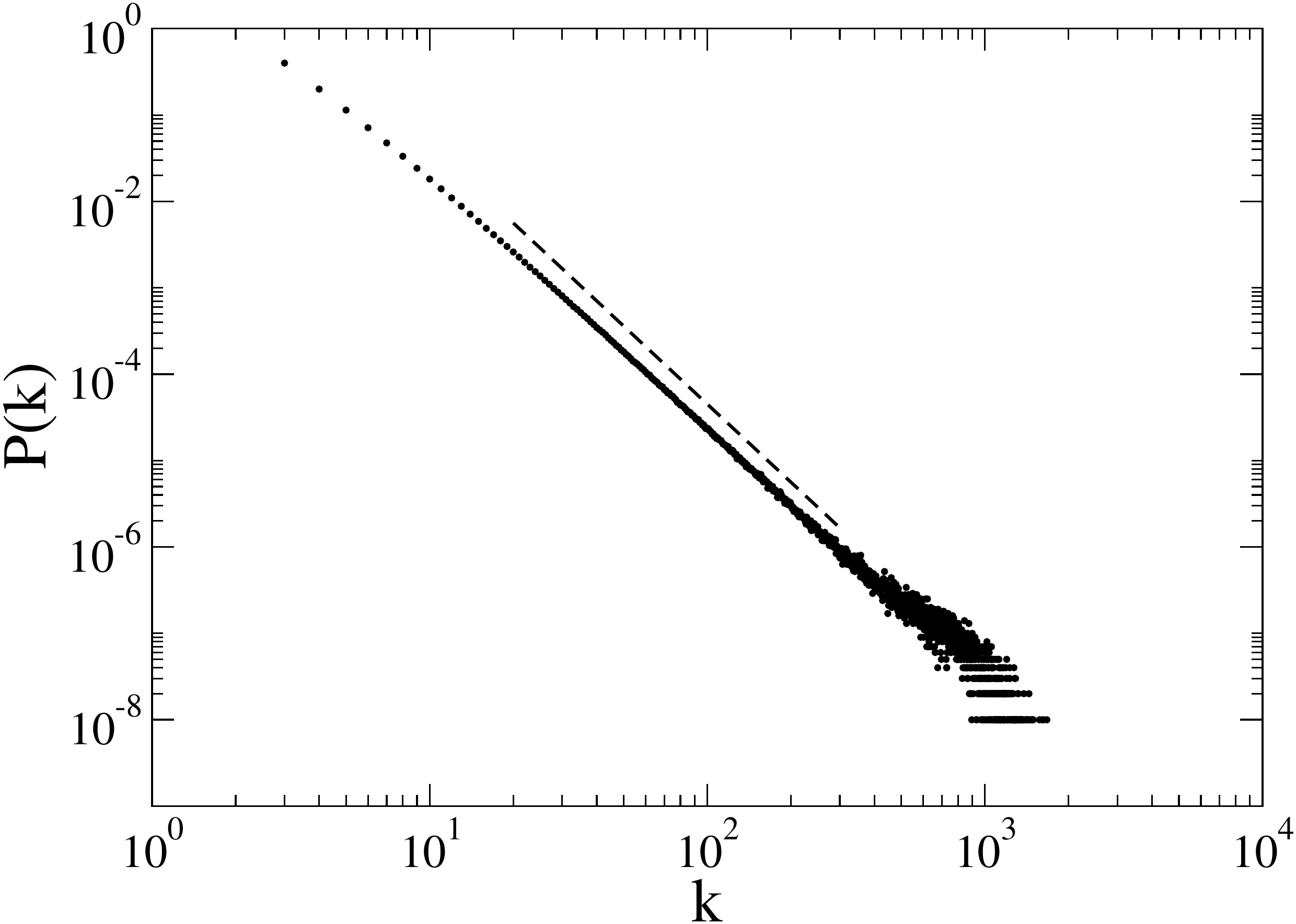}
  \caption{Distribution of the connectivity degree P(k) of the BA network. Dots are the average over $10^3$ networks of
  size $N=10^5$ and $m_0=3$. The dashed line has a slope $P(k) \sim k^{-3}$ and serves as a guide for the eyes.}
  \label{fig:NFS_distr}
 \end{figure}

These rules define what is know by \textit{Barabàsi-Albert (BA) model}, and generate a network with a  distribution of connectivity, say $P(k)$, that follows  a power-law degree distribution of the form $P(k) \sim k^{-\gamma}$, with $\gamma = 3.0$, in 
the thermodynamic limit, which is independent of the value of $m_0$, as shown in figure~\ref{fig:NFS_distr}.

 %\begin{equation}
 % P(k) \approx 2m_0^2 k^{-\gamma}
 % \label{eq:Distr_NFS}
 %\end{equation}

\begin{figure}[!htb]
 \centering
   \includegraphics[width=7cm]{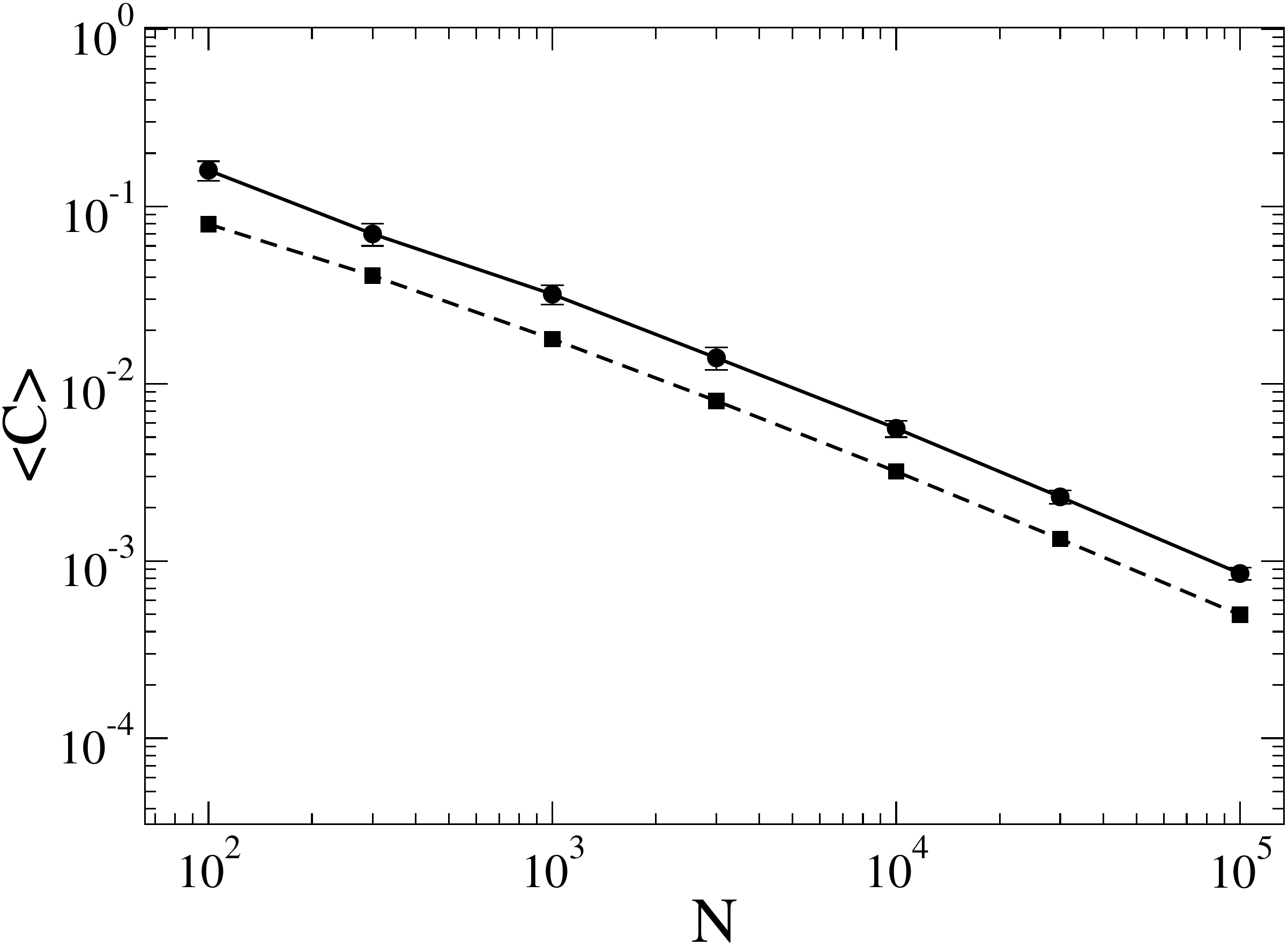}
   \caption{Clustering coefficient in function of the network size for BA network. The average was over 100 samples.
   The dots in the dashed line represents the theoretical value calculated from Eq.~\ref{eq:cluster_NFS} and the dots in the continuous line is obtained from simulations.}
  \label{fig:cluster_BA}
 \end{figure}

We can also calculate the clustering coefficient, say  $\langle C \rangle$, of the  BA network~\cite{dorogovtsev2004shortest}. It is the
 tendency of the network to form fully connected sub-graphs in the neighborhood of a given vertex, and 
 grows with the network size $N$ as:
  
 \begin{equation}
  \langle C \rangle \sim \frac{[\ln(N)]^2}{N}
  \label{eq:cluster_NFS}
 \end{equation} 
 We showed this behavior in figure \ref{fig:cluster_BA}. The simulation data follow the same bias as given by equation \ref{eq:cluster_NFS}.
 
Other important measure of networks is called shortest path length. The distance between
two any nodes $i$ and $j$ is defined as the number of links in
the shortest path that 
connects them, named $d_{ij}$.
The measure that represents the average over all shortest paths that link all the possible pairs of vertices in the  network is called the average shortest path length $\langle d \rangle$~\cite{dorogovtsev2004shortest}. For BA network, it is given by  $\langle d \rangle \sim \frac{\log N}{\log(\log N)}$, confirming its small world property~\cite{bollobas2003mathematical}.

Other feature that should be analysed is the degree correlation. The nodes of a network can present a tendency to connect 
with other nodes that have a similar or dissimilar degree. When the first case happens one says the network is
assortative correlated and if the second case occurs, the network is categorized as a disassortative correlated~\cite{associatividade}.
 
The simplest and most used way to quantify the degree correlation is given by the \textit{average degree of the nearest  neighbors} (nn) of a vertex $i$ with degree $k_{i}$~\cite{dorogovtsev2002evolution}, 
 \begin{equation} \label{corrum}
  k_{nn,i} = \frac{1}{k_{i}} \sum_{j \in \mathcal{N}(i)} k_{j},
 \end{equation}
 where the sum runs over by the nearest neighbors vertices of $i$, represented by the set $\mathcal{N}(i)$. The degree correlation 
 is obtained by the average degree of the nearest neighbors, $k_{nn}(k)$, for vertices of degree $k$ \cite{pastor2001dynamical}. That is,
 
 \begin{equation} \label{corrdois}
 k_{nn}(k) = \frac{1}{N_{k}} \sum_{i|k_{i} = k} k_{nn,i},
 \end{equation}
where $N_{k}$ is the number of nodes of degree $k$ and the sum runs over all vertices with the same degree $k$.
This quantity is related to the correlations between the degrees of connected nodes because in average it can be expressed as
 \begin{equation} \label{eq:corrtres}
  k_{nn}(k) = \sum_{k'} k'P(k'|k),
 \end{equation}
 where $P(k'|k)$ is the probability of a node with degree $k$ to have a neighbour node with degree $k'$.
 If degrees of neighboring vertices are uncorrelated, $P(k'|k)$ is just a function of $k'$
and $k_{nn} (k)$ is a constant. If $k_{nn}$ increases with $k$ then vertices with high degrees have a larger 
likelihood of being connected to each other. If $k_{nn}$ decreases with $k$, high degree vertices have larger 
probabilities of have neighbors with low degrees~\cite{pastor2001dynamical,barrat2008dynamical}. 

 \begin{figure}[!htb]
 \centering
   \includegraphics[width=7cm]{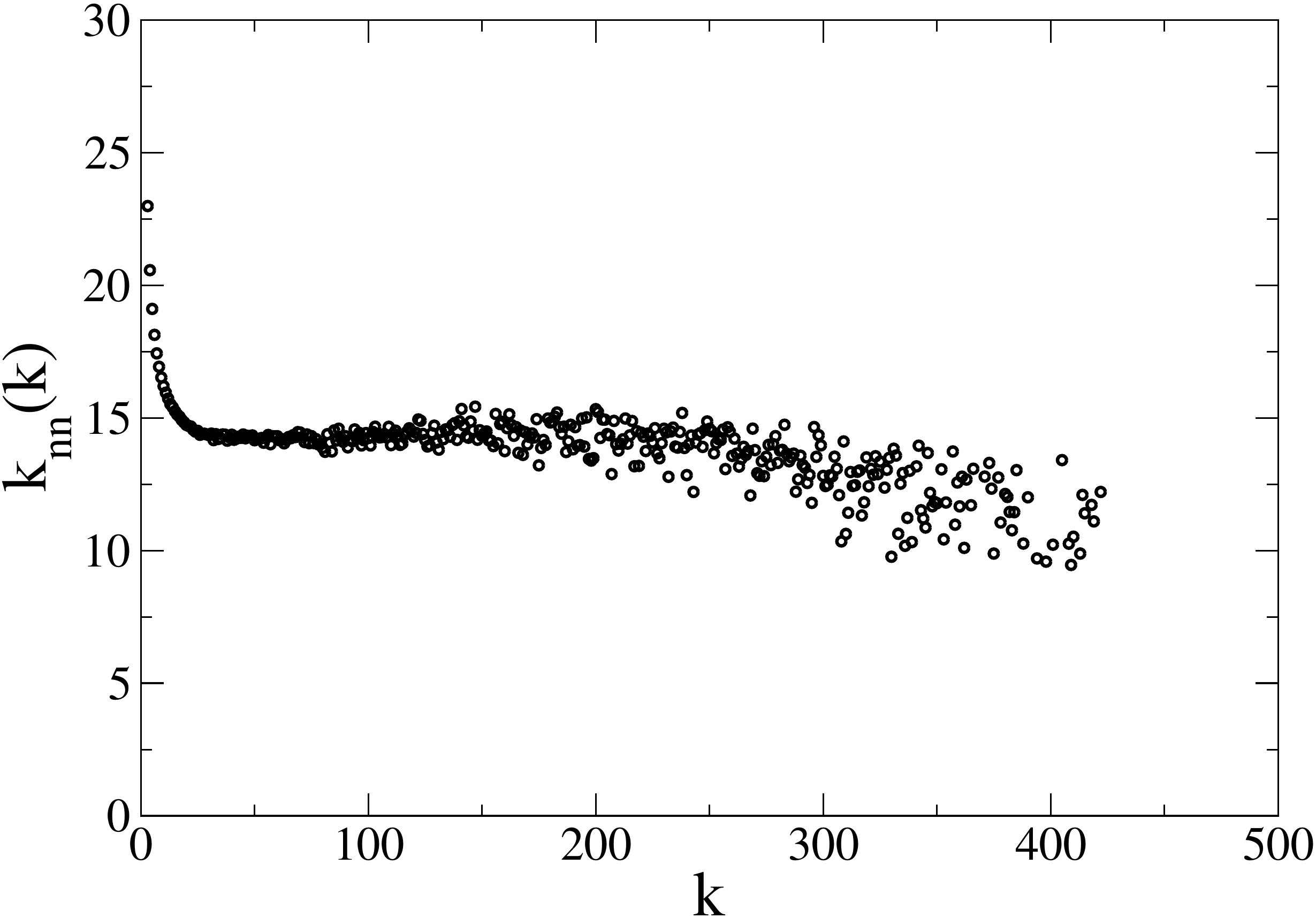}
   \caption{Degree correlation measured through the nearest-neighbors degree.
   It was used a BA network with size $N=10^4$,  and averaged over $10^3$ samples. 
  The preferential attachment rule of the BA networks affects just the connectivity of nodes recently added in the network, the ones with small $k$. They connect primarily with hubs, creating a disassortative correlation 
for small values of $k$. As long as the degree grows, the network becomes almost uncorrelated.}
  \label{fig:knnBA}
 \end{figure} 
 
 \begin{table*}[!htb]
	\caption{Table with all main informations of the networks that were investigated in this work.
	The mean clustering coefficient $\langle C \rangle$ and the Pearson correlation coefficient $c_P$ are obtained for a sample of 1000 networks with size $N=10^{4}$. The average shortest path length $\langle d\rangle$ is obtained for a sample of at least 20 networks with the same size. For networks with Euclidean distance we consider always $\alpha_A =3$, since the topological phase transition occurs for $\alpha_A \approx 2.$ The preferential attachment rule is shown in the column $\prod(k_i|j)$. P(k) is the degree distribution form of each model, and $\gamma$ is the exponent related to a power law degree distribution that characterizes the first three networks that were investigated.
}
	\label{tab:table_general}       % Give a unique label
	\centering
	\begin{tabular}{lllllll}
		\hline\noalign{\smallskip}
		Network & ~ $\Pi(k_i|j)$  &    $P(k)$ & ~$\gamma$  &
		~$\langle C \rangle$  &  ~$\langle d \rangle$ &$c_{p}$ \\
		\hline\noalign{\smallskip}
	Barabási-Albert &  $\frac{k_i}{\sum_n k_n}$ &   $\sim k^{-\gamma}$ &  3  & 0.0055(5) & 4.3(1) &  -0.037(4) \\
			\noalign{\smallskip}\hline\noalign{\smallskip}
		Fitness Model & $\frac{\eta_i k_i}{\sum_n k_n}$ & $\sim k^{-\gamma}$ &  2.25  & 0.028(7) & 4.1(2) & -0.09(1)\\
		(Bianconi-Barabási)  & & & & & & \\			
		\noalign{\smallskip}\hline\noalign{\smallskip}
		Homophilic Model & $\frac{(1- {\cal A}_{ij})k_i}{\sum_n(1- {\cal A}_{in}) k_n}$ & $\sim k^{-\gamma}$ &  2.75 & 0.015(3) & 4.2(2) & -0.038(4) \\
				\noalign{\smallskip}\hline\noalign{\smallskip}
	Euclidean Distance Model &   $\frac{k_i r_{ij}^{-\alpha_A} }{\sum_n k_i r_{in}^{-\alpha_A}   }$ &  $\sim e_q^{-k/\kappa}$ & - &  0.0019(2) & 4.7(1) & 0.034(7) \\
				\noalign{\smallskip}\hline\noalign{\smallskip}
			Fitness Model &   $\frac{ \eta_i k_i r_{ij}^{-\alpha_A} }{\sum_n \eta_n k_i r_{in}^{-\alpha_A}   }$ &   $e_q^{-k/\kappa}$ & - & 0.0034(4) & 4.6(1) & -0.046(8) \\
			with euclidean distance & & & & & & \\	
		\noalign{\smallskip}\hline\noalign{\smallskip}
			Homophilic Model &   $\frac{ (1- {\cal A}_{ij})  k_i r_{ij}^{-\alpha_A} }{\sum_n (1- {\cal A}_{in}) k_i r_{in}^{-\alpha_A}   }$ &   $e_q^{-k/\kappa}$  & - &
			 0.0020(2) & 4.7(1) & 0.028(7) \\
		with euclidean distance & & & & & \\
	%	\noalign{\smallskip}\hline\noalign{\smallskip}
	%		Phone calls  & ?& ?& &? & ? & \\		
			%Collaboration   & & & & & \\	
			%Email  & & & & & \\	
			
		\noalign{\smallskip
		}\hline
	\end{tabular}
	% Or use
\end{table*}

The BA network is weakly disassortative as we showed in the figure~\ref{fig:knnBA}. We observe that the 
preferential attachment interferes just in the connectivity of nodes recently added in the network. According 
to the rule of the model, these nodes connect preferably with hubs, creating a disassortative correlation 
for small values of $k$. But, as long as the degree grows, the network becomes almost uncorrelated.

We also can use the Pearson coefficient, named $c_P$, to quantify degree correlations, according to the expression~\cite{barrat2008dynamical}:
 
 \begin{equation}
  c_P = \frac{\sum_e j_e k_e/E - [\sum_e (j_e+k_e)/(2E)]^2}{[\sum_e (j_e^2+k_e^2)/(2E)] - [\sum_e (j_e+k_e)/(2E)]^2},
  \label{eq:pearson}
 \end{equation}
where $j_e$ and $k_e$ are the degrees of the nodes that are in the beginning and in the end of the edge $e$,  and $E$ 
is the total number of connections. This quantity ranges from $-1$ to $1$ meaning disassortative and assortative networks, 
respectively.
It is a complementary information to the $k_{nn}(k)$ measure. While the latter provides how the degree correlation can vary with $k$, the Pearson coefficient ($c_P$) quantifies the degree correlation of the entire network according to a scale ranging from -1 to 1.
This measure was also used to complement the characterization of 
a topological phase transition on growth and preferential attachment model that consider the euclidean 
distance between the nodes, as we will see in section~\ref{sec:metrica}. In addition, it will be useful to compare the synthetic networks with real ones, in section~\ref{sec:realnets}.

All the main information of the BA network is summarized in the table~\ref{tab:table_general}, as well the information about other networks that were also treated in this paper.
In general, real networks present a power law degree distribution with $2<\gamma<3$~\cite{nature_communications,associatividade}.
So, the BA model is restricted to describe a large set of them because its degree exponent is fixed $\gamma \approx 3$. 
Next, we show other features that can be added to the model to make it more realistic.

\subsection{Fitness Model: Bianconi-Barabási Network}
\label{sec:BB}
The original BA model produces a power-law network with the presence of sites that become privileged that is, 
with more connections over time. But this model does not taking into account the competitiveness, this means, 
the ability of younger nodes to acquire new neighbors.
Facebook, for example, has become one of the most
visited sites in a short period of time when compared to the Google, an older search website. Another example
is the growth of corporations where some newer ones concentrate more services than older ones. We can simulate this situation 
including a intrinsic characteristic in each node, called fitness. In social networks, fitness would represent 
an individual's attribute of becoming more popular due to some quality of him/her. % that sets he/she apart from most. 
In networks, it represent the probability of a node to become a hub quickly. 

This characteristic was observed in real networks by Bianconi and Barabàsi, who later proposed
an alternative model including the fitness factor $\eta_i$ of each node $i$~\cite{bianconi2001competition}. The
algorithm is similar to the BA network, but each site
connects to an existing node on the network with a probability that, in addition to depending on $k$ connectivity, 
is also proportional to the attractiveness $\eta_i$. The choice of $\eta_i \in [0,1]$ is usually given by a uniform 
distribution $\rho(\eta_i)$~\cite{bianconi2001competition}, and the connection probability is defined by:

\begin{equation}
  \Pi(k_i|j)=\frac{\eta_i k_i}{\sum_n \eta_n k_n}.
  \label{eq:Bianc_Barabasi_Prob}
 \end{equation}

\begin{figure}[!htb]
 \centering
    \includegraphics[width=7cm]{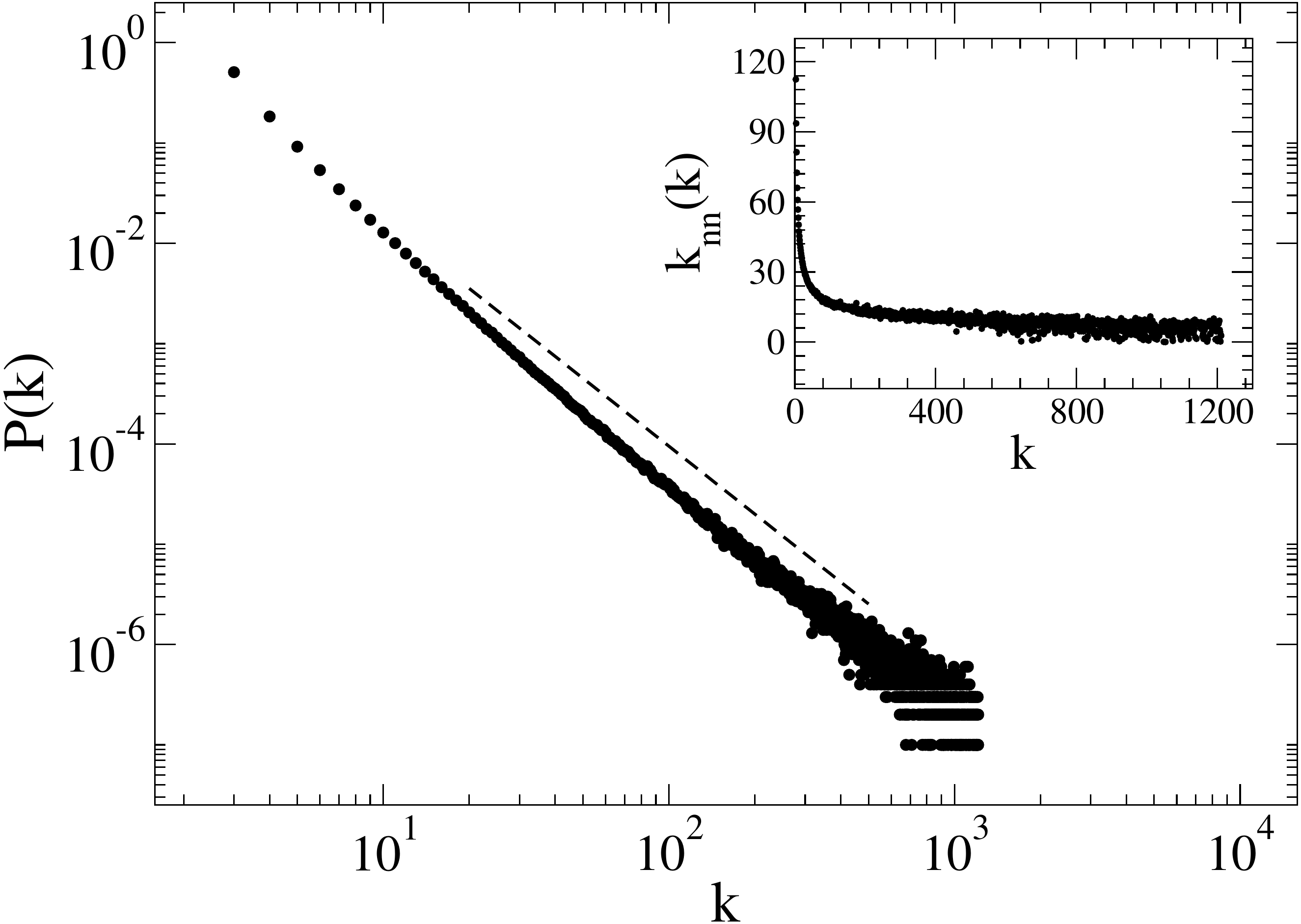}
 \caption{ Distribution of connectivity for Bianconi-Barabàsi model with $m_0 = 3,  N = 10^4$
  based on 1000 network realizations. The graph is on the log-log scale. The dashed line, with slope $P(k) \sim k^{-2.25}$,
  is a guide for the eyes.
This network has more privileged sites and, consequently, more hubs than the BA network, which explains its smaller value of $\gamma$.
  Inset: Degree correlation measured through the nearest-neighbors degree for the same set of networks. The behavior is similar to that one we found for BA network.}
 \label{fig:Distr_BBA}
\end{figure}

When the fitness parameter is imposed, the network remains a power-law degree distribution but with an exponent 
less than 3 (see figure \ref{fig:Distr_BBA}). According to the literature, $\gamma = 2.25$ in the thermodynamic limit. There are more privileged sites and, consequently, more hubs than the BA network, which makes the $\gamma$ exponent smaller, that is, the network is more heterogeneous. In the inset of figure \ref{fig:Distr_BBA} we show the degree correlation measured through the nearest-neighbors degree for the Fitness model. The behavior is similar to that one we found for BA network. We also calculate the mean clustering coefficient $\langle C \rangle$, the average shortest path length $\langle d \rangle$, and the Pearson correlation coefficient $c_P$. These informations are shown in table~\ref{tab:table_general}. We observed that, when compared to BA model, this network presents a higher cluster coefficient and a lower Pearson correlation coefficient, but the average shortest path length pretty does not change. 

\subsection{Homophilic Model}
\label{sec:Afinidade}

In a social network, people tend to relate to others who share common characteristics such as musical taste,
football team, religion, and work. To take this tendency into account in social network models, we can include a 
connection parameter called homophily.

Almeida and colleagues~\cite{homophilic} proposed a model introducing this parameter through an intrinsic property
value of each node, called $\eta_i$, similar to the previous model. The homophily between any two nodes $i$ and $j$, say  
$\mathcal{A}_{ij}$, is defined
as the module of the difference between $\eta_i$ and $\eta_j$, that is, $\mathcal{A}_{ij} = |\eta_i -\eta_j|$. 
The lower is $\mathcal{A}_{ij}$, the greater the affinity between both and, consequently, the greater the probability to 
connect with each other. The proposed algorithm is as follows:

\begin{itemize}
\item It starts with $m_0$ sites connected to each other, in the same way as in the BA network, but introducing a 
characteristic $\eta_i$ for each node, chosen randomly in a uniform distribution $p(\eta)$ in the interval $[0,1]$.

\item For each time step, add a node $j$ that links to other $m_0$ nodes already on the network. Each $j$ node connects
preferably to a node $i$ according to the probability

\begin{equation}
\Pi(k_i|j)=\frac{(1-\mathcal{A}_{ij})k_i}{\sum_n (1-\mathcal{A}_{in}) k_n}
\label{eq:Prob_Afini}
\end{equation}
\item The procedure of the second item is repeated until the network reaches a previously established size $N$.
\end{itemize}

\begin{figure}[!htb]
\centering
    \includegraphics[width=7cm]{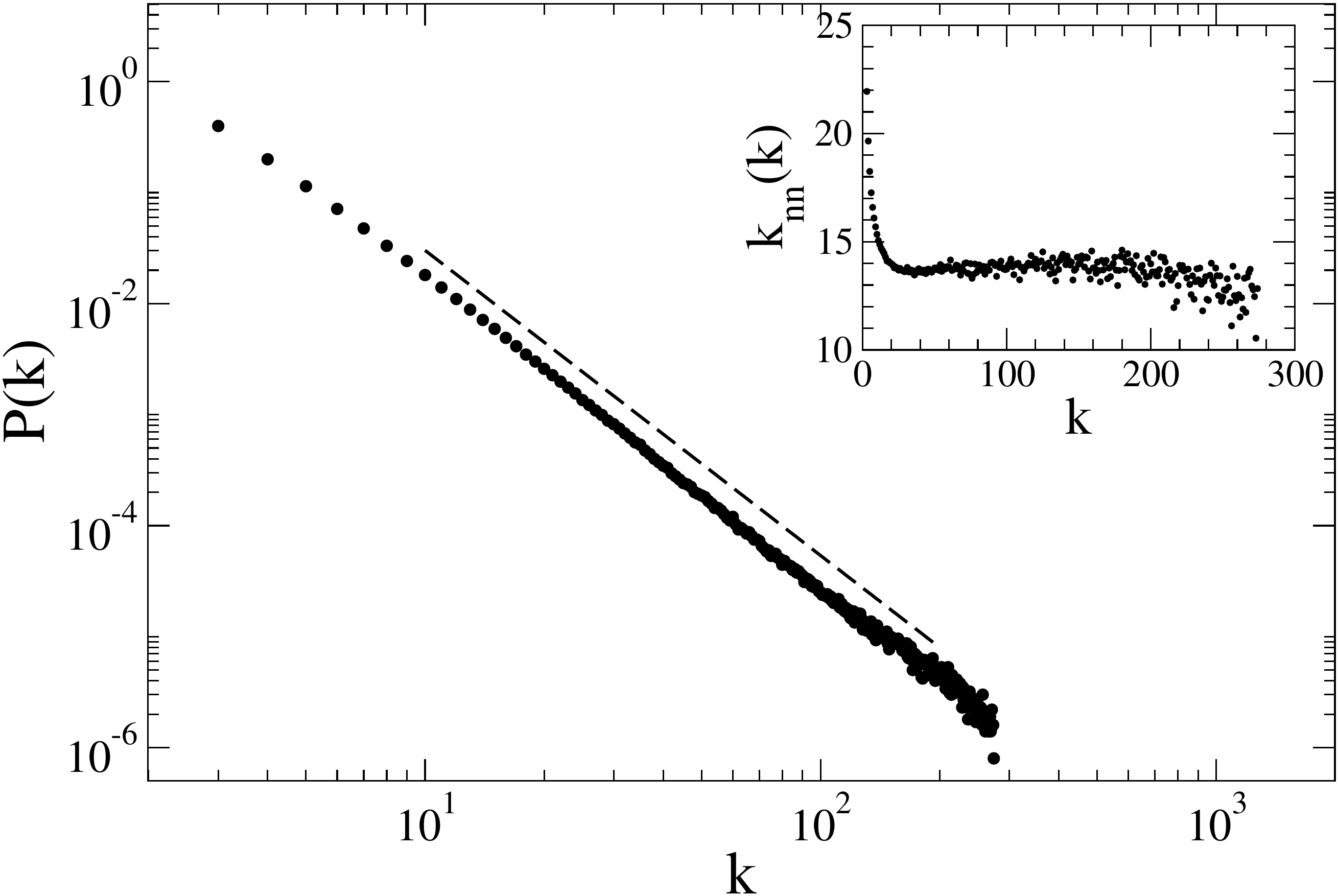}
\caption{Distribution of connectivity for Homophilic model network. It was used networks with size $N =10^4$ 
and $m_0 =3 $, based on $1000$ network realizations. The dashed line has slope $P(k)\sim k^{-2.75}$
as a guide for the eyes.
Its $\gamma = 2.75$ is smaller than the $\gamma = 3.0$ for BA model and greater than $\gamma = 2.25$, obtained in the Fitness netwok. This happen because the democratization of which node can become a hub is not as prominent as in the Fitness network, as we discussed in the text. Inset: Degree correlation measured through the nearest-neighbors degree for the same set of networks. The behavior is similar to that one we found for BA and Fitness networks.}
\label{fig:Distr_afinidade}
\end{figure}

The competition between the degree of connectivity and the affinity between the nodes generates a network with a power law 
degree distribution, but with $\gamma \sim 2.75$, as we shown in figure~\ref{fig:Distr_afinidade}. This value is lesser than the exponent obtained in the BA model 
($\gamma = 3.0$) but it is greater than the value obtained in the Fitness network ($\gamma = 2.25$).
This difference is explained because in the BA network, only the degree of connectivity is considered to make links, which generates the
phenomenon ``rich gets richer". In the Fitness model, nodes newly inserted in the network can become hubs more 
easily as long as they have high fitness. That is, there is a democratization because a node can become hub regardless of 
its age,  as is the case of facebook and google network, for example. In the homophilic model, a node $j$ can assume a value of $\eta_j = 0.5$, for example. When this happens, if it tries
to connect to a node $i$ that has $\eta_i = 0.3$ or another node $k$  which has $\eta_k = 0.7$, the affinities between 
both pairs $ij$ and $jk$ are the same. So, in this case, according to the expression (\ref{eq:Prob_Afini}) who dictates the
preference in the connection is the degree of the node~\cite{homophilic}.

 In the inset of figure \ref{fig:Distr_afinidade} we show the degree correlation measured through the nearest-neighbors degree for the Homophilic model. The behavior is similar to that one we found for the other networks. We also calculate the mean clustering coefficient $\langle C \rangle$, the average shortest path length $\langle d \rangle$, and the Pearson correlation coefficient $c_P$. These informations are shown in table~\ref{tab:table_general}. We observed that, when compared to BA model, this network presents almost the same  $\langle d \rangle$ and $c_P$ but a slightly higher clustering coefficient. 

\subsection{Networks with Euclidean distance}
\label{sec:metrica}

The models presented above do not take into account the spatial distance between the agents that compose the network. But 
in many real systems, that variable plays an important role. For example, in the city model proposed by
Ribeiro and colaborators \cite{ribeiro2017model}, the authors observed how the Euclidean distance influences the
potential of cities and in scale's law to measure socio-economic and infrastructure indicators.  
There are other works that showed the relation between social interaction and spatial properties~\cite{Laniado2017,Liu2019,Lengyel2015GeographiesOA}. For example, in the paper \cite{Laniado2017}, the authors analyzed online social networks and they found that spatial distance restricts who interacts with whom and denser connected
groups tend to arise at shorter spatial distances.
In the following subsections we reconstructed the standard models shown previously including euclidean distance between the nodes as an attachment ingredient.

\subsubsection{Model proposed by Soares {\it et. al}}

 \begin{figure}[!htbt]
 \centering
  \includegraphics[width=7cm]{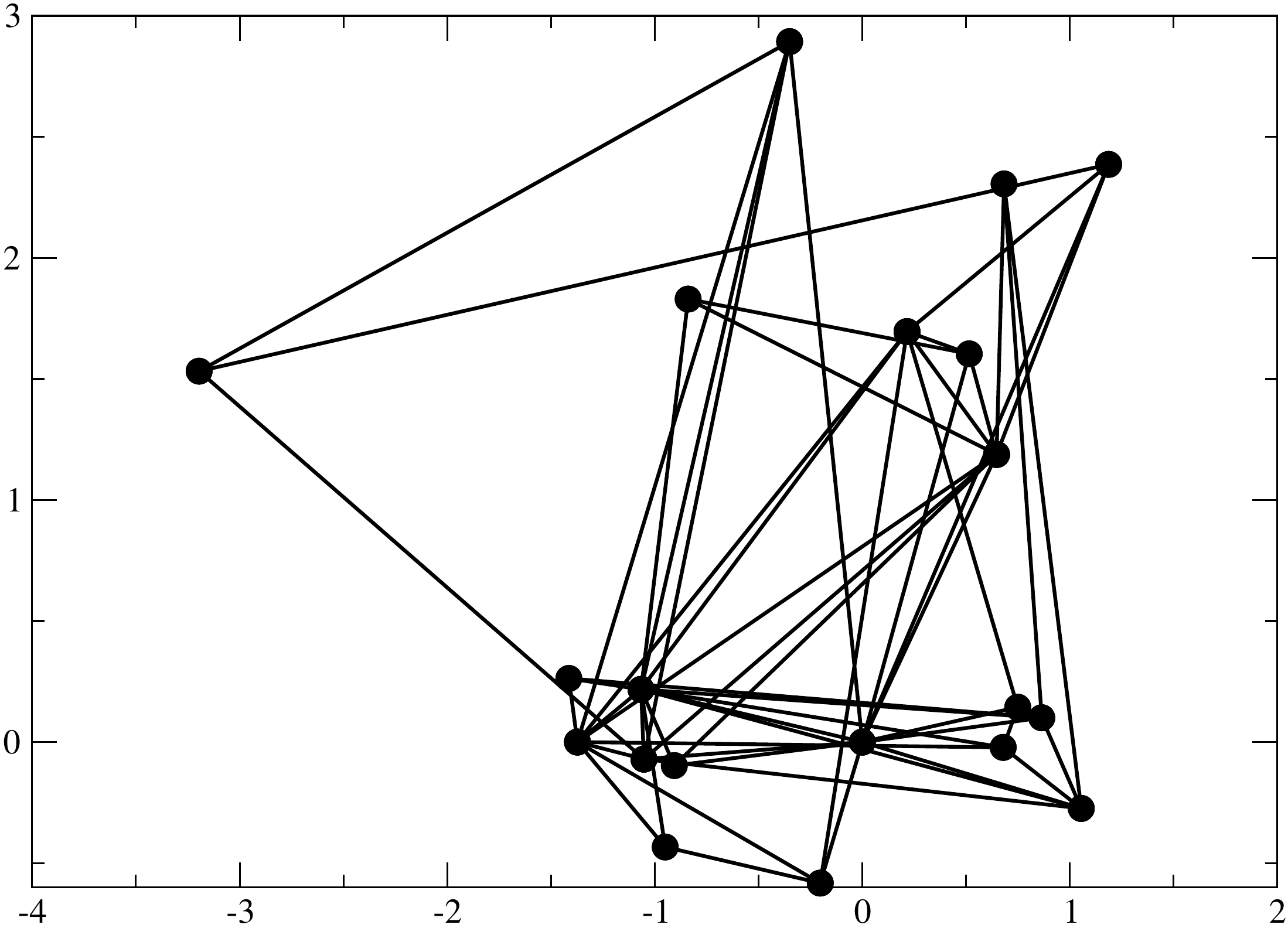}
  \caption{An example of a network with size $N = 20$ generated according to the algorithm proposed by Soares 
  {\it et. al.}~\cite{soares2005preferential} using $\alpha_A = 2$ and $\alpha_G = 2$. Note that the 
  nearest links are more likely but long-range connections can also happen.}
  \label{fig:rededistancia}
 \end{figure}
 
\begin{figure*}[!htb]
\centering
    \includegraphics[width=7.5cm]{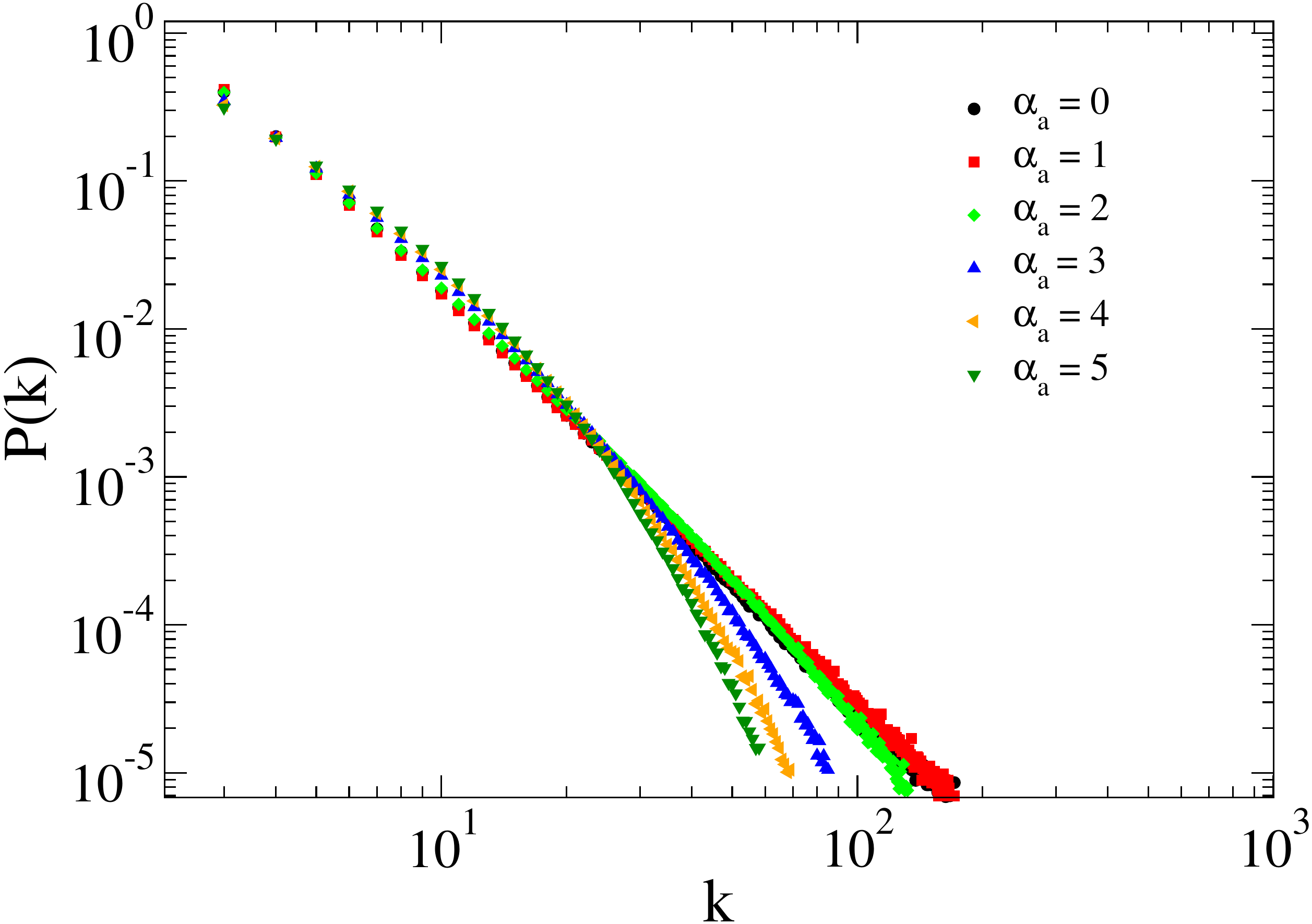}\quad
    \includegraphics[width=7.5cm]{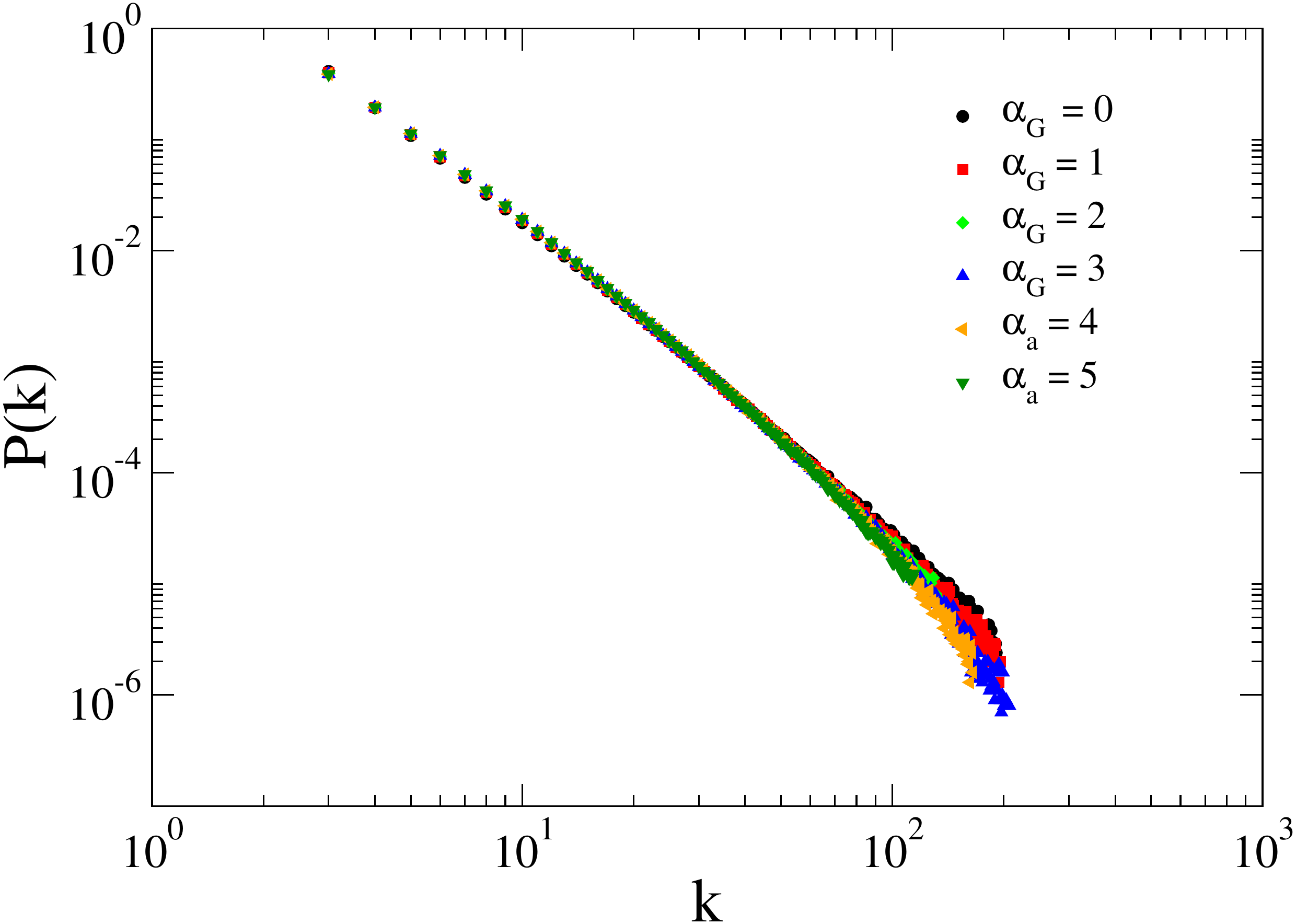}
    
    ~~~~~~~~~(a) ~~~~~~~~~~~~~~~~~~~~~~~~~~~~~~~~~~~~~~~~~~~~~~~~~~~~~~~~~~~~~(b)
\caption{
(a) Distribution of connectivity for different values of $\alpha_A$ and a fixed $\alpha_G = 2$.
By varying $\alpha_A$, a topological transition appears  close to $\alpha_A = 2$. 
 For $\alpha_A = 0$, the BA network is recovered, meaning that spatial distance between nodes is not taken into account.
 The network changes from  a completely heterogeneous network ($\alpha_A = 0$) to an increasingly homogeneous network as $\alpha_A$ tends to infinity.  
(b)  Distribution of connectivity for  different values of $\alpha_G$ and a fixed $\alpha_A = 2 $ (right). By varying $\alpha_G$ the distribution does not change significantly.
Results obtained with networks of size $N = 10^4$,  averaged over 1000 samples.}
\label{fig:Distr_Natal}
\end{figure*}

Soares and colleagues~\cite{soares2005preferential} built a model in which the preferred connection dynamics 
happens according to the degree of connectivity but also considers the Euclidean distance between the nodes.
To build the model, we consider that each site is inserted on a continuous plane. The first node is added at an arbitrary
distance from the origin and the others are isotropically distributed with a probability
$P_G(r) \propto r^{-(2+\alpha_G)}$, which depends on the distance $r$ from the center of mass, which is positioned at $r_{cm}$ from the origin and is re-calculate 
at each time step. The exponent $\alpha_G$ ($G$ refers to the word growth) is responsible for the network growth, that is, 
defines how close or distant the nodes will be placed. The calculation of the position of the center of mass is given by 
$\mathbf{r}_{cm} = \frac{1}{M} \sum_{n=1}^N m_n \mathbf{r}_n$ where $m_n$ is the mass of the node $n$, and $\mathbf{r}_n$ is the vector-distance of this node to the  origin, and $M = \sum_{n = 1}^N m_n$ is the total mass. The network has a total of $N$ nodes, and we can consider each node 
with mass $m_n = 1$, so we have $\mathbf{r}_{cm} = \frac{1}{N} \sum_{n = 1}^N \mathbf{r}_n$. Each new site $j$ connects to a pre-existing node $i$ 
following a rule of preferential connection that depends on the distance between them, $r_{ij}$ and the degree of connectivity of the node $i$, that is,

\begin{equation}
\Pi(k_i|j)=\frac{k_i r_{ij}^{-\alpha_A}}{\sum_n k_n r_{in}^{-\alpha_A}}.
  \label{eq:ProbConectNatal}
\end{equation}

The $\alpha_A$ exponent ($A$ refers to the word attachment) controls the influence of spatial distance between sites in the
preferential attachment. If $\alpha_A = 0$, we recover the BA network that does not take into account the spatial distance
between the nodes. This model preserves the preferential attachment according to the degree of the nodes, but also taking into 
account  the geographical distance as a criterion to dispute for links. In figure \ref{fig:rededistancia} we show 
a plot of an example of a network generated according to this algorithm.

%The model is built according to the following algorithm:
%\begin{itemize}
%\item A site $j = 1$ is added at an arbitrary distance from the origin on the continuous plane.
%\item A second site $i = 2$ is placed at a random distance $r$ from the site according to the distribution
%$P_G(r) \propto r^{-(2+ \alpha_G)}$.
%\item As new nodes are inserted, the center of mass is calculated and the node is added to the plane at
%distance $r$ from the center of mass according to the distribution $P_G(r)$. Each inserted node preferably connects to 
%nodes already present in the network according to the expression:
%\begin{equation}
% P_A \propto \frac{k_i}{r_i^{\alpha_A}},
%\end{equation}
%where $\alpha_A \geq 0$ characterizes the attachment pattern of the network, therefore, sub-index $A$.

%\item The previous procedures are repeated until the network reaches an established size $N$.
%\end{itemize}

\begin{figure*}[!htb]
\centering
    \includegraphics[width=7.5cm]{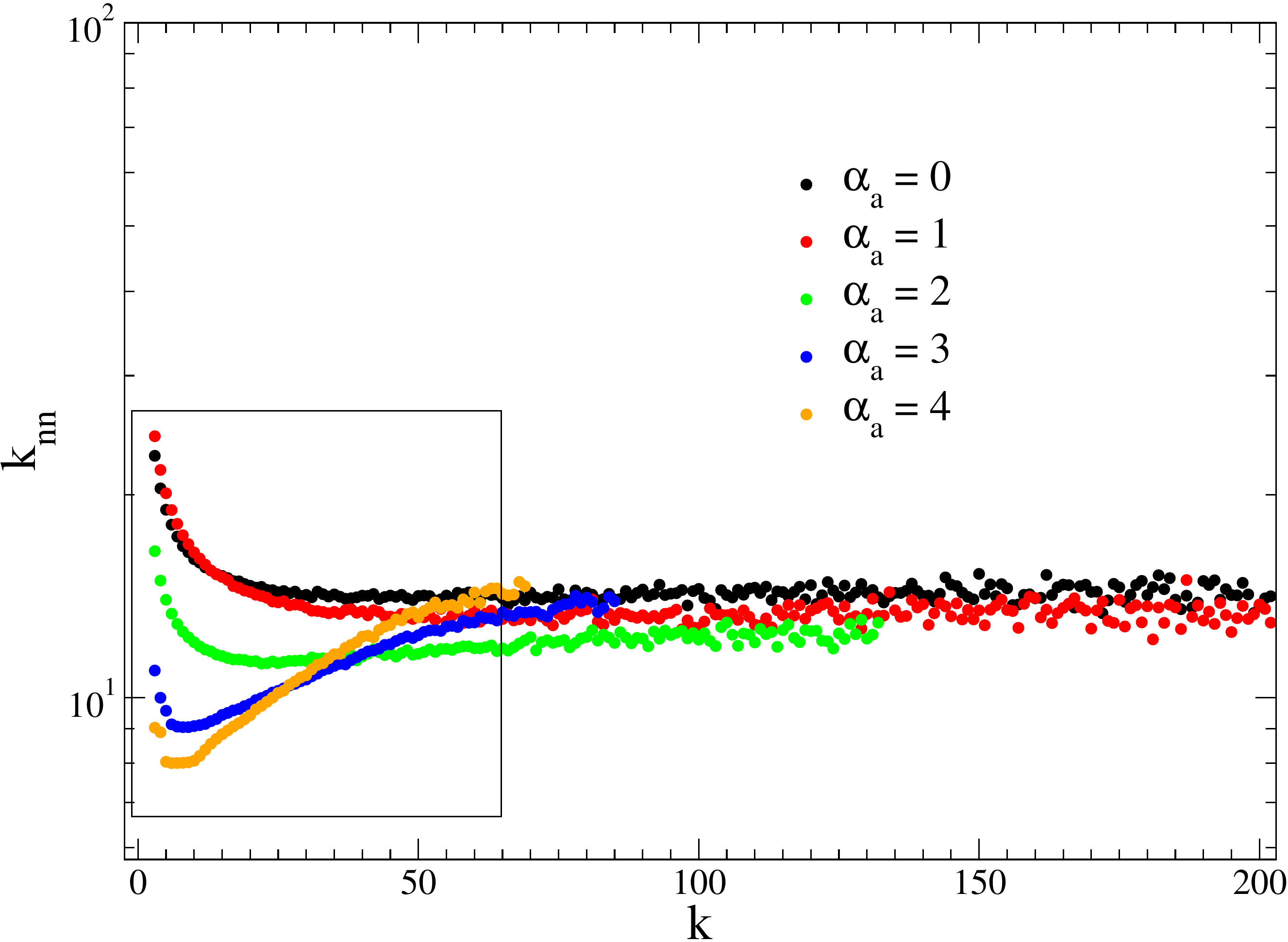}\quad
    \includegraphics[width=7.5cm]{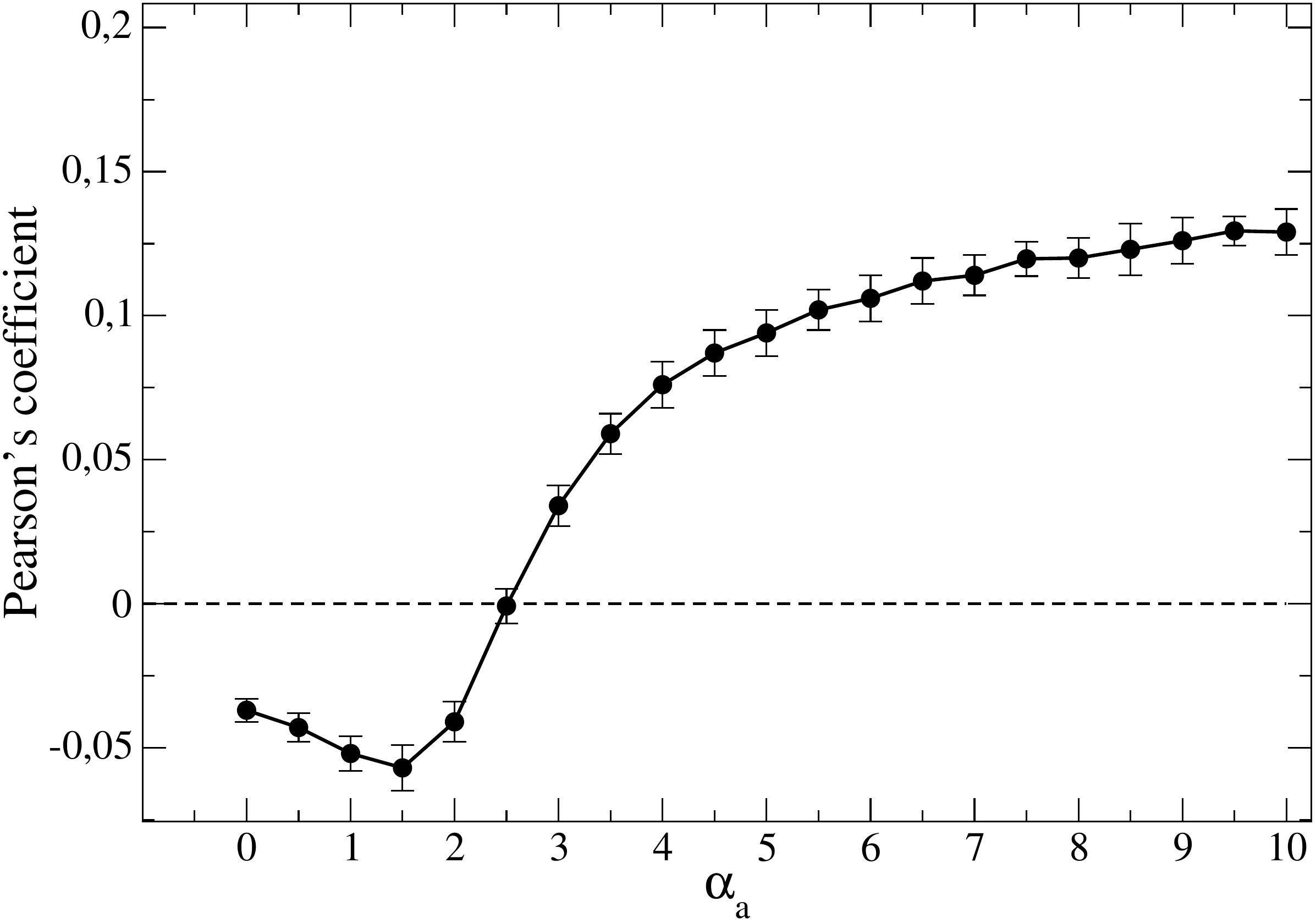}
    
        ~~~~~~~~~(a) ~~~~~~~~~~~~~~~~~~~~~~~~~~~~~~~~~~~~~~~~~~~~~~~~~~~~~~~~~~~~~(b)
\caption{Degree correlation for a network of size $N = 10^4$ and different values of $\alpha_A$. In figure (a) we show the
measure of $k_{nn}$ in function of $k$. Highlighted the change from a weakly disassociative regime to an associative one. 
In figure (b),the Pearson's coefficient as a function of $\alpha_A$. 
Close to $\alpha_a = 2$, Pearson's coefficient changes from a negative value, 
which characterizes a disassociative network, to a positive value, which characterizes an associative network. In both (a) and (b) the average was made over 1000 samples.}
\label{fig:knn_alfaA}
\end{figure*}

%If we take $\alpha_A = 0$, we return to the well-known Barabasi-Albert model. 

\begin{figure}[!htb]
\centering
    \includegraphics[width=7cm]{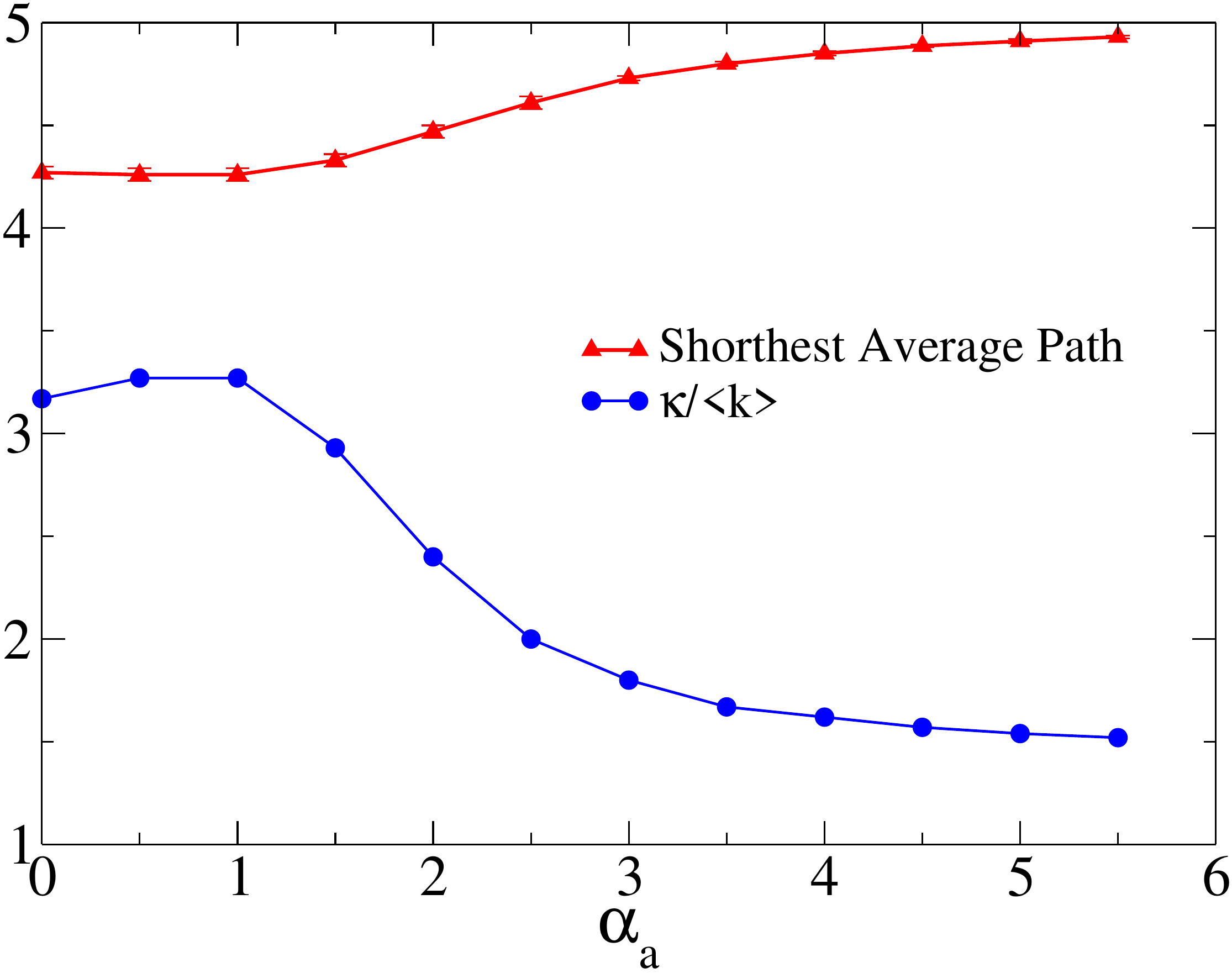}
\caption{Measure of network heterogeneity defined by $\kappa/\langle k \rangle$ (blue line). If $\kappa/\langle k \rangle > 1$ the network is considered
heterogeneous, otherwise it is homogeneous. We note that the network becomes more homogeneous as $\alpha_A$ increases. We also 
plot the calculation of the shortest average path (red line). When the network becomes more homogeneous, 
the shortest average path increases, as the hubs disappear and then the distance (number of links that connects any  pair of nodes)
between the nodes increases. This measure also reinforces the topological phase transition. We performed $1000$ samples of
networks with size $N = 10^4$.}
\label{fig:kappa_alfaA}
\end{figure}

Numerical results show that the parameter $\alpha_G$ does not affect the behavior of the connectivity distribution
$P(k)$ of the network (see Figure \ref{fig:Distr_Natal}(b)). This parameter refers just to the distance distribution in relation
to the center of mass, and  acts only on size scale  but not on the structure of the network, and consequently  it  does not impact on the preferential attachment rules. On the other hand, as $\alpha_A$ increases, 
the connectivity distribution changes (see figure \ref{fig:Distr_Natal}(a)).
Soares {\it et. al.}~\cite{soares2005preferential} showed that the degree distributions of networks generated according to their model are very well fitted with the form
\begin{equation}
  P(k)=P(0) e_q^{-k/k_0}
  \label{eq:DistrTsallis}
\end{equation}
where $k_0>0$ is the characteristic number of connections, $P(0)$ is a constant to be normalized, $q$ is the entropic index
and $e_q^x$ is the  $q$-\textit{exponential} defined by

\begin{eqnarray}
e_q^x \equiv \left[1+(1-q)x\right]^{1/(1-q)}, 
\label{eq:qExp}
\end{eqnarray}
where the natural exponential function is a particular case: $e^x =  e_{q=1}^x$.

The authors~\cite{soares2005preferential} showed that both $k_0$ and $q$ are functions of $\alpha_A$. So, as $\alpha_A$ increases, a topological phase transition occurs in the connectivity 
distribution~\cite{nunes2017role,soares2005preferential}. The network changes from 
a completely heterogeneous network ($\alpha_A = 0$) to an increasingly homogeneous network as $\alpha_A$ tends to infinity.
Such phase transition also appears in the degree correlation of the nodes, as we show in the calculation of $k_{nn}(k)$ 
for different values of $\alpha_A$ (see figure \ref{fig:knn_alfaA}(a)).
The transition is clearer in the graph \ref{fig:knn_alfaA}(b) in which we show the calculation of Pearson's coefficient 
as a function of $\alpha_a$. Close to $\alpha_a = 2$, Pearson's coefficient changes from a negative value, 
which characterizes a disassociative network, to a positive value, which characterizes an associative network.

Finally, other two more evidence that the topological phase transiton can be discussed.
We can measure the level of heterogeneity of a network using the quantity $\kappa = \langle k^2 \rangle / \langle k \rangle$, 
where $\langle k^p \rangle$ is the $p-th$ moment of the degree distribution. If $\kappa/\langle k \rangle > 1$ 
the network is considered heterogeneous because it means that the second moment of the degree distribution can diverge when $N \rightarrow \infty$ while 
for homogeneous networks $\kappa/\langle k \rangle \approx 1 $~\cite{barrat2008dynamical}. As can be seen in 
figure~\ref{fig:kappa_alfaA}, the network becomes more homogeneous as $\alpha_A$ increases because $\kappa/\langle k \rangle$ 
decreases and approaches to one. We also calculated the average shortest path length. When the network becomes more homogeneous, 
the average shortest path length increases because the number of  hubs decreases and consequently the path between the nodes increases. 
This measure, also shown in figure~\ref{fig:kappa_alfaA}, reinforces the topological phase transition.

\subsubsection{Variations of the model proposed by Soares {\it et. al.}}

It is possible to include euclidian distance in the homophilic and the fitness models, as investigated by Nunes and collaborators~\cite{nunes2017role}. For example, when we study the social interaciton in a
city~\cite{ribeiro2017model}, the parameter $\eta_i$ can represent the influence of different
places localized in the city. So it is possible to use the fitness model with Euclidean distance to 
try to explain, for example, why some places in a city is more attractiveness than others to open a store, coffee shop or 
gas station. We also can use the homophilic model including spatial distance to study the influence of the 
topology in a formation of neighborhoods, since 
people tend to cluster with people that have a similar social class, religion or workplace~\cite{bisgin2010investigating,complexity2017,entropy2018}.
  
The algorithms used to construct both models were already shown in previous sections. Now, we just have to include 
the metric, using the function $P_G \sim r^{-(\alpha_G+2)}$ to distribute the nodes in a continuous plane and change the 
preferential attachment rules that become,

 \begin{equation}
  \Pi(k_i|j)=\frac{\eta_i k_i r_{ij}^{-\alpha_A}}{\sum_n \eta_n k_n r_{in}^{-\alpha_A}} ~~~~~~~\text{and} 
  \label{eq:Distr_Metrics}
 \end{equation}

 \begin{equation}
  \Pi(k_i|j)=\frac{(1-\mathcal{A}_{ij})k_i r_{ij}^{-\alpha_A}}{\sum_n (1-\mathcal{A}_{in}) k_n r_{in}^{-\alpha_A}},
 \end{equation}
for fitness and homophilic models, respectively.
Nunes~\cite{nunes2017role} also shown a topological phase transition, 
as $\alpha_A$ increases for fitness model and no influence of the parameter $\alpha_G$ in the pattern of the connectivity 
distribution. We obtained the same results for homophilic networks. The data are not shown because they are very similar 
to the results shown in figure \ref{fig:Distr_Natal}.

\section{Real Networks}
\label{sec:realnets}

Most of social networks are assortative while technological ones tend to be more disassociative~\cite{associatividade,
Newman}. To support this evidence and show that the models studied in this article are successful in modeling real systems, 
we have chosen three real networks to investigated two distinct properties: degree distribution and 
assortativity. The networks are:

\begin{itemize}
 \item Phone Calls: nodes represent cell phone users and the edges exist if they 
 have called each other at least once during the investigated period. Data are from~\cite{phonecalls}.
 
 \item Collaboration network: each node represents an author in a scientific collaboration and 
 the edges between them represent a co-authored at least one paper in the period from January 1993 
 to April 2003. The data are obtained from arXiv preprint  Condense Matter Physics~\cite{collaboration}.
 
 \item Email:  nodes are email adress and a directed link from one node to another represent at least one email
 sent. The data are collected during 112 days in the University of Kiel (Germany)~\cite{email}.
\end{itemize}

According to the table \ref{tab:table1}, the Pearson's coefficient of phone calls and collaboration networks are 
positive while for email network this coefficient is negative. The first two are social networks and they are basically
related to family/friendship and professional interactions, respectively. 
In reference~\cite{associatividade}, 
Newman found similar results for biology and mathematics coauthorship. However, the email network, 
although it also describes some social interaction, behaves more as a technological network.
In the reference cited above, the author also found similar value for World-Wide-Web.

\begin{table}[!htb]
\caption{Size $N$ and Pearson's correlation coefficient $r$ for different real networks. We compared the values 
with the Pearson's coefficient calculated for synthetic networks with the same size. For the phone calls network, 
we used the homophilic network including euclidean distance ($\alpha_A = 5$). For the collaboration network, 
we used the BA network including euclidean distance ($\alpha_A = 5$) and finally for the email network, 
we used the fitness network including euclidean distance ($\alpha_A = 1$).  }
\label{tab:table1}       % Give a unique label
\begin{tabular}{llll}
\hline\noalign{\smallskip}
Real Network & ~ $N$ & ~ $c_P$ & $c_P$ (synthetic network) \\
\noalign{\smallskip}\hline\noalign{\smallskip}
Phone Calls & 36594 & 0.282 & ~~~~~~~~ 0.120\\
Collaboration & 23132 & 0.134 & ~~~~~~~~ 0.112\\
Email & 57194 & -0.075 & ~~~~~~~ - 0.078\\
\noalign{\smallskip
}\hline
\end{tabular}
% Or use
\end{table}
Now, we can compare this real systems with our investigated models. 
In the case of phone calls network, we used the homophilic model and we investigated how the euclidean distance 
between the nodes of the system affects the network's degree distribution. Homophilic model was chosen 
because it is reasonable to assume that telephone calls happen between people who have a certain affinity with
each other, whether for personal, family or professional reasons. This hypothesis is corroborated in recent 
works~\cite{currarini2016simple,bisgin2010investigating}. 

\begin{figure}[!htb]
\centering
    \includegraphics[width=8.5cm]{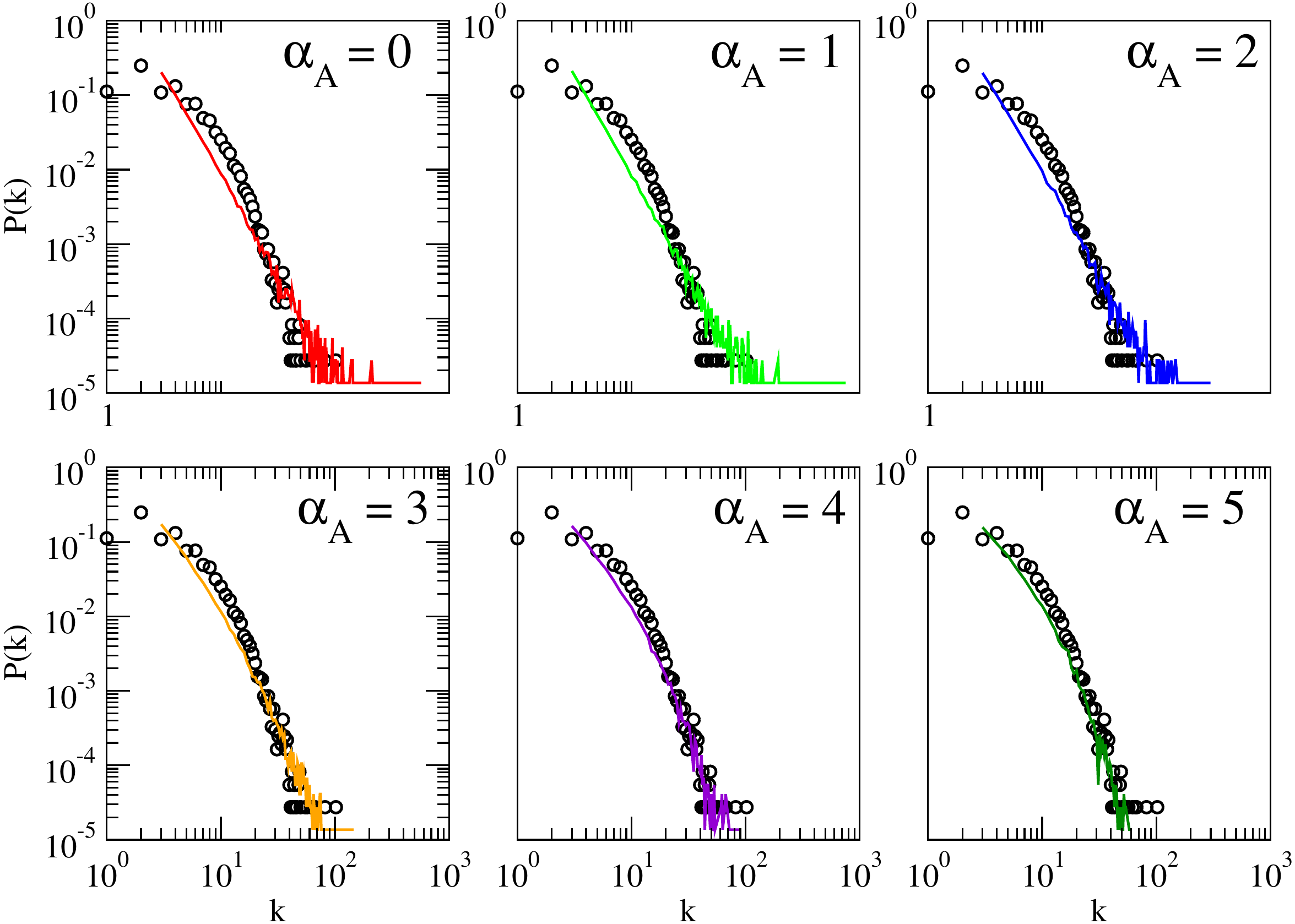}
\caption{Degree  distribution of a phone calls network compared with the distinct degree distributions of synthetic networks 
with the same size generated according to the homophilic model including Euclidean distance. 
Black points represent real network data and solid colored lines are related to 
synthetic networks.
The synthetic network with $\alpha_A=5$ fits better the data.
}
\label{fig:phonecalls}
\end{figure}

The Pearson correlation coefficient of the investigated synthetic network is not very similar to the value obtained for the 
real network (see table \ref{tab:table1}). However we appreciated how accurate the degree distribution of this synthetic network is when compared to 
the real one. As shown in figure \ref{fig:phonecalls}, when we used the attachment parameter $\alpha_A = 5$, the fits 
works extremely well, emphasizing the importance of considering geographic distance between the elements of the system
when modeling real social networks. Indeed, a lot of work have followed this 
line~\cite{Lengyel2015GeographiesOA,Laniado2017,Liu2019}.

\begin{figure}[!htb]
\centering
    \includegraphics[width=8.5cm]{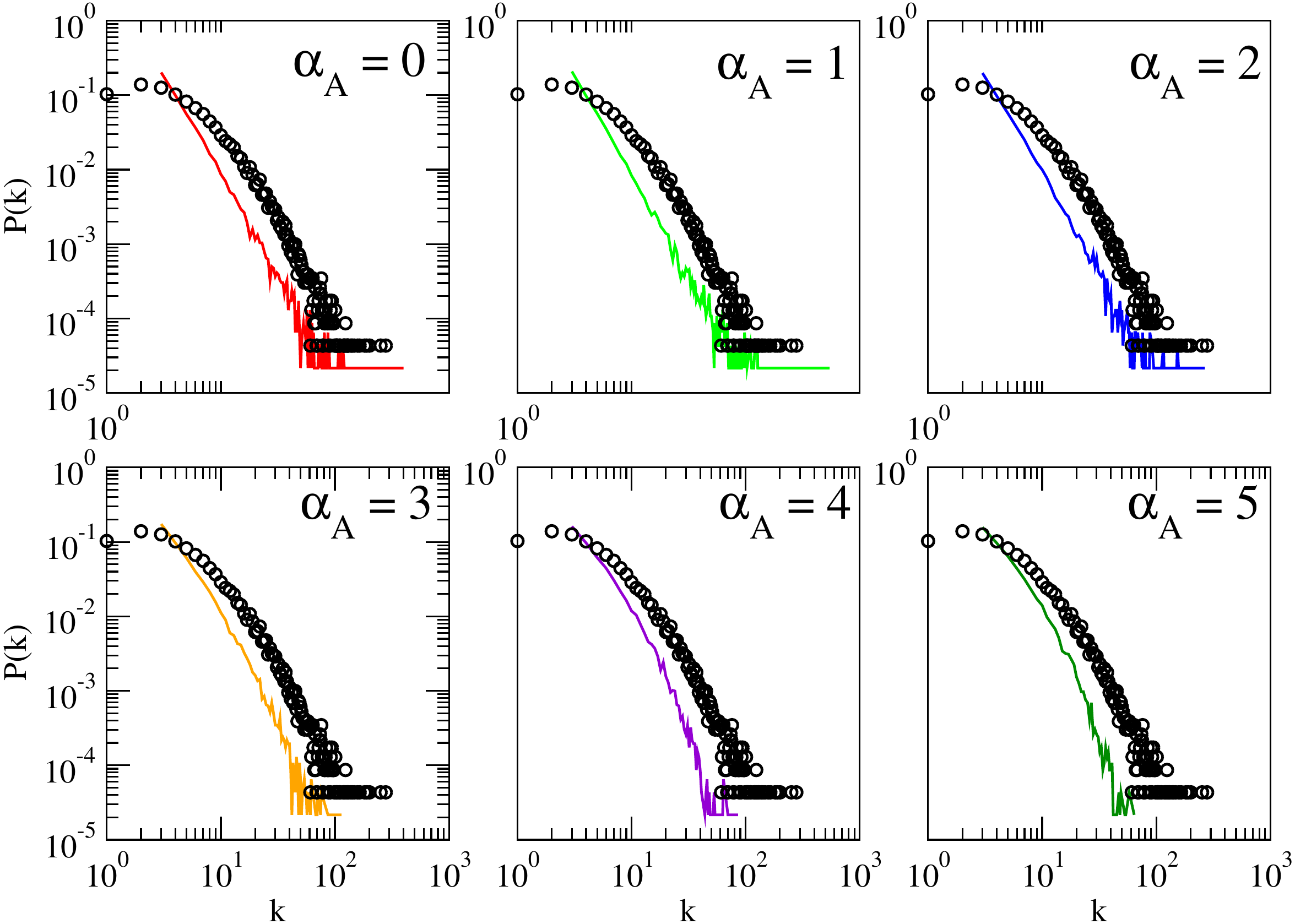}
\caption{Degree  distribution of a collaboration network compared with the distinct degree distributions of synthetic networks 
with the same size generated according to the Barab\`{a}si-Albert model including Euclidean distance.  We observe that 
the last scenario ($\alpha_A=5$) fits better the real data. Black points represent real network data and solid colored lines are related to 
synthetic networks.}
\label{fig:collaboration}
\end{figure}

The same analysis can be done for the collaboration network. The Pearson correlation coefficient 
of the synthetic network is similar to the one calculated for the real system. In this study, the 
only change was the synthetic network investigated. We chosed the traditional Barab\`{a}si-Albert model 
but also including the Euclidean distance and, as we showed in figure \ref{fig:collaboration}, the 
fit using $\alpha_A = 5$ is accurate as well. In networks of scientific co-authorship more distinguished researchers, 
such as university professors, tend to publish works with less famous researchers, such as their graduate students.
This support both assumptions: the BA preferential attachment rule according to the degree of the node 
and the influence of the distance between the elements of the system. For the collaboration network, the fitness and homophilic models
also showed reasonable results. As long as we increased the value of $\alpha_A$, the preferential connection rule involving 
the Euclidean distance prevails in relation to the others.

Finally, the email network presents a very similar Pearson correlation coefficient with compared 
to the fitness synthetic network considering the euclidean distante. But here the parameter $\alpha_A = 1$ fits 
better the degree distribution of real data. It shows a smaller impact of the geographical distance of the nodes in 
technological networks than in social ones. This can also be related to the fact that this real network has directed
links. As this email network was obtained from a university, the fitness model was chosen based on the fact that,
in academia, students tend to send more emails to teachers than the otherwise. So the message sent depends on how 
influential (greater fitness) the reciever is.

\begin{figure}[!htb]
\centering
    \includegraphics[width=8.5cm]{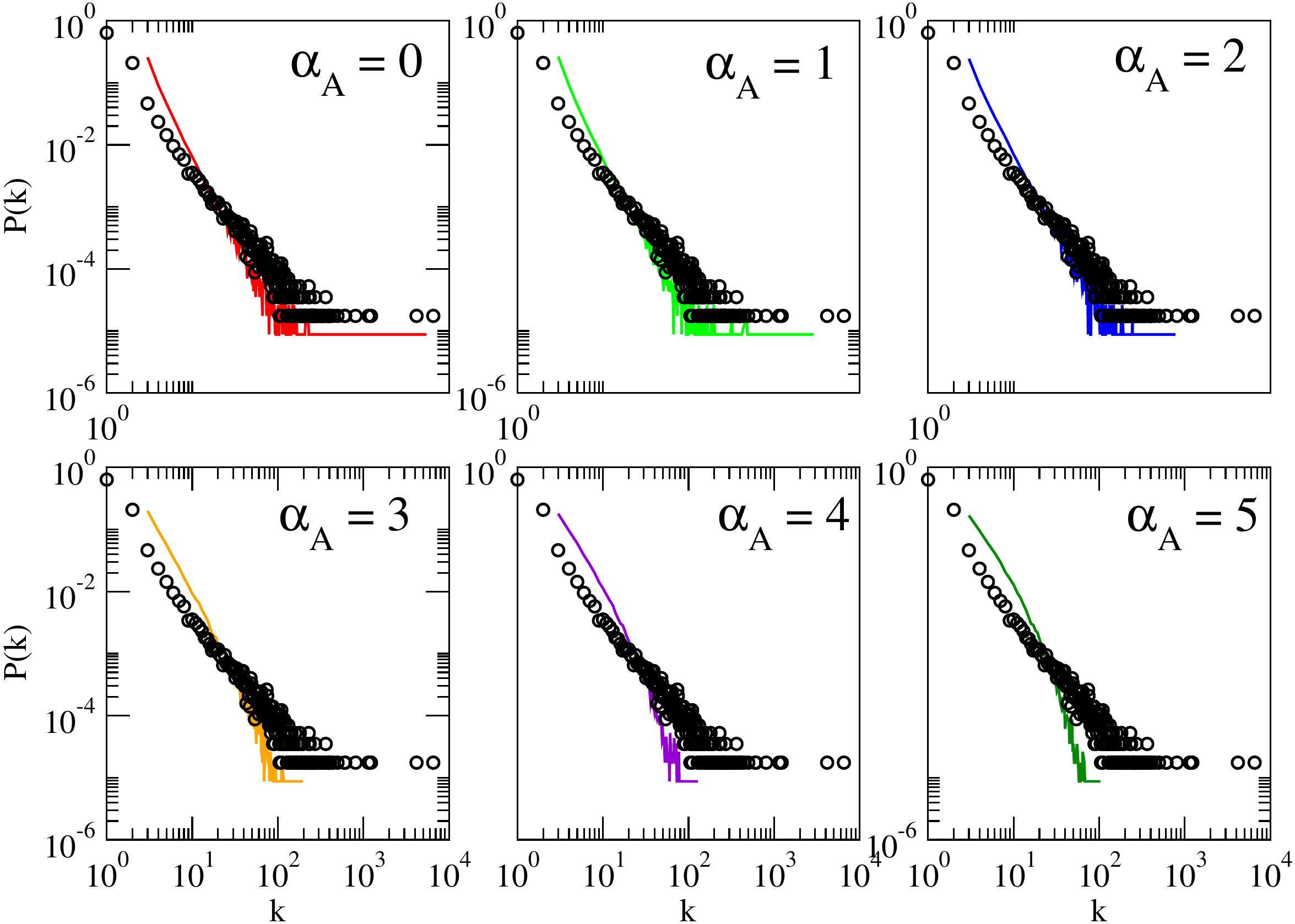}
\caption{Degree  distribution of an email network compared with the distinct degree distributions of synthetic networks 
with the same size generated according to the fitness model including Euclidean distance.  We observe that 
the second scenario ($\alpha_A=1$) fits better the real data. Black  points represent real network data and solid colored lines are related to 
synthetic networks.}
\label{fig:email}
\end{figure}

\section{Conclusions}

In this work we have studied network models with growth and different rules of preferential attachment. 
We reviewed some important algorithms such as the Barabasi-Albert model and others that includes fitness, 
homophily and/or Euclidean distance as strategies to make connections between nodes. From an applicable 
perspective, these models are useful to model real-world networks because they present 
characteristics found in social sytems and also in technological ones, as we showed in the last section.

Our results corroborated with evidences 
that power-law degree structured is not very common in real systems. We evaluated two social and one technological network and we compared the degree distribution of these networks to degree distributions generated by growth and preferential attachment models. Our main conclusion is that the  real networks analysed are better fitted with models which consider traits as fitness, homophily and euclidean distance between nodes. We observed that geographic distance between nodes seems to be an important factor to model specially real social systems.
This feature changes the form of the degree distribution of a power law to a q-exponential according to the model proposed by Soares and collaborators~\cite{soares2005preferential}.
Our results are in agreement with recent studies involving real networks~\cite{nature_communications,broido2019scale,holme2019,phonecalls,email,collaboration,ribeiro2017,Laniado2017,Liu2019,Lengyel2015GeographiesOA}.

We also supplemented the characterization of these synthetic networks investigating measures as clustering, average shortest path length, degree distribution 
and assortativity.
% Some of them seem to be a relevant feature to differentiate social networks from technological ones.

Finally, it is important to mention that many dynamical processes as epidemics, rumor propagation and 
synchronization were extensively investigated in scale-free to- pologies as the Barabàsi-Albert network. However
the study of these dynamics in substrates where the distance between the elements of the system is taken into account 
needs to further advance, since this element has already been shown to be very important. Even on online social networks~\cite{Lengyel2015GeographiesOA,currarini2016simple,bisgin2010investigating,Laniado2017,Liu2018,Liu2019}, it seems to influence the connection between the nodes, as well as fitness and homophily.

\label{sec:conclu}

\section*{Acknowledgments}
This work was partially supported by the Brazilian agencies CAPES, CNPq and FAPEMIG.
The authors acknowledges computational time at DFI-UFLA.  Ang\'{e}lica S. Mata thanks the support from FAPEMIG (Grant No. APQ-02482-18) and CNPq (Grant No.  423185/2018-7).
Fabiano L. Ribeiro thanks the support from  CNPq (405921/2016-0) and CAPES (88881.119533/2016-01).

%\bibliographystyle{apsrev}
%\bibliography{refbib,fabiano.bib}

\end{document}